\documentclass[conference,a4paper]{IEEEtran}

\usepackage[utf8]{inputenc} 
\usepackage[T1]{fontenc}
\usepackage{url}             
\usepackage{cite}            

\usepackage[cmex10]{amsmath}  
\usepackage{mleftright}       
\mleftright                   

\usepackage{graphicx}         
\usepackage{booktabs}         

\usepackage{graphics}
\usepackage{scalefnt,setspace}
\usepackage{epsf}
\usepackage{calc}
\usepackage{graphicx}
\usepackage{psfrag}
\usepackage{flushend}

\usepackage{hyperref}

\usepackage{tikz}
\usetikzlibrary{arrows,decorations.pathmorphing,positioning,backgrounds,fit,calc,shapes.geometric}

\begin{document}

\def\naive{na\"{\i}ve}
\newtheorem{problem}{Problem}
\newtheorem{definition}{Definition}
\newtheorem{lemma}{Lemma}
\newtheorem{proposition}{Proposition}
\newtheorem{corollary}{Corollary}
\newtheorem{example}{Example}
\newtheorem{conjecture}{Conjecture}
\newtheorem{algorithm}{Algorithm}
\newtheorem{theorem}{Theorem}
\newtheorem{exercise}{Exercise}
\newtheorem{remark}{Remark}

\newcommand{\bml}[1]{\mbox{\boldmath $ #1 $}}
\newcommand{\ssbml}[1]{\scriptsize{\mbox{\boldmath $ #1 $}}}
\newcommand{\sbml}[1]{\mbox{\scriptsize{\boldmath $ #1 $}}}
\newcommand{\hathat}[1]{\mbox{$\hat{\hat{#1}}$}}

\newcommand{\olsi}[1]{\,\overline{\!{#1}}} 
\newcommand{\ols}[1]{\mskip.5\thinmuskip\overline{\mskip-.5\thinmuskip {#1} \mskip-.5\thinmuskip}\mskip.5\thinmuskip} 
\newcommand{\ulsi}[1]{\!\underline{\,{#1}}} 
\newcommand{\uls}[1]{\mskip.5\thinmuskip\underline{\mskip-.5\thinmuskip {#1} \mskip-.5\thinmuskip}\mskip.5\thinmuskip} 

\newcommand{\be}{\begin{equation}}
\newcommand{\ee}{\end{equation}}
\newcommand{\bea}{\begin{eqnarray}}
\newcommand{\eea}{\end{eqnarray}}
\newcommand{\beaa}{\begin{eqnarray*}}
\newcommand{\eeaa}{\end{eqnarray*}}
\newcommand{\ben}{\begin{enumerate}}
\newcommand{\een}{\end{enumerate}}
\newcommand{\bi}{\begin{itemize}}
\newcommand{\ei}{\end{itemize}}
\newcommand{\eqd}{\stackrel{\Delta}{=}}
\newcommand{\limn}{\lim_{n\rightarrow\infty}}

\newcommand{\one}{\frac{1}{n}}
\newcommand{\half}{\frac{1}{2}}
\newcommand{\reals}{{\rm I\!R}}
\newcommand{\onei}{{\rm 1\!\!\!\:I}}
\newcommand{\NN}{{\rm I\!\!\!\;N}}
\newcommand{\EE}{{\rm I\!\!\!\;E}}
\newcommand{\Var}{\mathrm{Var}}
\newcommand{\Cov}{\mathrm{Cov}}

\renewcommand{\IEEEQED}{\IEEEQEDopen}

\newcommand{\markov}{\mbox{$-\hspace{-2.3mm}\circ\hspace{1mm}$}}
\def\squarebox#1{\hbox to #1{\hfill\vbox to #1{\vfill}}}
\newcommand{\qed}{\hspace*{\fill}
          \vbox{\hrule\hbox{\vrule\squarebox{.667em}\vrule}\hrule}\smallskip}

\title{Relay Channels with Unreliable Helpers}

\author{%
  \IEEEauthorblockN{Yossef Steinberg}
  \IEEEauthorblockA{%
    Technion - IIT\\
    Haifa 3200003, ISRAEL\\
    {\tt ysteinbe@technion.ac.il}}
}

\maketitle

\begin{abstract}
The relay channel with unreliable helper is introduced an studied.
The model is that of a classical relay channel where the input from the relay
to the channel has an extra primitive link
 whose presence is not assured a priori. The extra link represents a helper
 who may decide not to cooperate in transmission.
The goal is to devise robust coding schemes that exploit all the relay links
when they are present, 
but can also operate,
possibly at reduced rates, when the extra primitive link (helper) is absent.
The capacity region of this class of problems is defined, and fully characterized
for degraded relay channels. 
The degraded Gaussian relay channel with unreliable relay link is solved.
\end{abstract}


\section{Introduction}
\label{sec:introduction}
The\makeatletter{\renewcommand*{\@makefnmark}{}
\footnotetext{
Part of this work was presented at the 2023 International Symposium on Information Theory, 
June 25-30 2023, Taipei, Taiwan.

This research was supported by the 
ISRAEL SCIENCE FOUNDATION (grant No. 2589/20).}\makeatother} 
relay channel (RC), introduced by van der Meulen~\cite{vanderMeulen:71p}, 
is one of the cornerstones of network information theory. 
In its basic form, it is the simplest (and first) model that describes
cooperation between receivers in a communication system, involving a main transmitter,
main receiver, and a relay node (transmitter+receiver).
The relay does not have messages of its own to send or decode. Its only task
is to help the main receiver to decode the messages intended to him, 
by transmitting a signal that is based on the output it receives, in a strictly causal manner.
A multi-letter characterization of the RC capacity was derived in~\cite{vanderMeulen:71p}.
Cover and El Gamal in~\cite{CoverElGamal:79p} derived bounds on the RC capacity, and characterized
it for  the degraded and reversely degraded models. 
A comprehensive overview of models and results on the RC
can be found in~\cite{ElGamalKim:11b}.

Cooperation between users as means to enhance performance of communication
networks has been a subject of intensive research in the last decades. 
The literature on these topics is vast, thus only a brief overview
is presented here, to put things in context. 
Willems introduced in~\cite{Willems:83p} the multiple access channel (MAC) with partially cooperating encoders
and derived its capacity region.
The cooperation takes place over limited capacity links, that are independent of the main channel.
Such links are often termed as conference links. 
In~\cite{WillemsVanderMeulen:85p} Willems and van~der~Meulen suggested the model of MAC
with cribbing encoders, and derived its capacity region for all forms of cribbing. 
Dabora and Servetto~\cite{DaboraServetto:06p} studied the broadcast channel (BC)
with cooperating decoders, and characterized its capacity for degraded BC. 
The decoders cooperate via limited capacity links that are independent of the
main BC, as in Willem's model~\cite{Willems:83p}. 
Independently, Liang and Veeravalli~\cite{LiangVeeravalli:07p}, \cite{Liang:05z} 
introduced the Relay Broadcast Channel (RBC), where the decoders serve as relays for each other.
This can be viewed also as an extension of  the basic RC of~\cite{vanderMeulen:71p}
to the BC, with various types of relays: cooperative relaying of one or two users, and dedicated relays.
In  particular, the results in~\cite{DaboraServetto:06p} can be obtained as special cases of the RBC
with one sided primitive relay.

The coding schemes developed in these works depend on the existence of the corresponding cooperation
resources - independent links, relays, sideway channels and more, that sometimes are used also for network management
and are not secured for cooperation only.
In modern ad hoc wireless communication networks, these resources can be fortuitous users, possibly from
a neighboring network, who may or may not agree to
serve as helpers in the system. 
Therefore their availability is not guaranteed a priori, even in cases where the statistics of  the main channel is perfectly known.
 Moreover, in certain situations
the users in  the system cannot be informed whether or not the helpers are available, 
and whether they participate in transmission.
Hence there is a need to devise robust coding schemes, that can exploit the cooperation resources when they are available,
but can still operate, possibly at reduced rates, when they are absent. 
The physically degraded BC with unreliable cooperation links was introduced in~\cite{Steinberg:14c},
and its capacity region fully characterized.
The MAC with unreliable cribbing encoders was also suggested in~\cite{Steinberg:14c}, and bounds were derived
on its capacity region. 
The works~\cite{HuleihelSteinberg:16c,HuleihelSteinberg:17p,ItzhakSteinberg:21a} 
further develop and sharpen the results of~\cite{Steinberg:14c}.

The purpose of this paper is to extend these ideas to the relay channel. 
We introduce a relay channel as in~\cite{vanderMeulen:71p} and~\cite{CoverElGamal:79p}, but with an
additional primitive link - representing the helper - whose signal is not guaranteed to arrive to the decoder.
Its (operational) capacity region is defined as the set of all rate pairs
that are achievable with one coding scheme, where the encoder and relay do not know a priori
whether the helper is active. The capacity region of this channel is characterized for the degraded RC,
and the Gaussian example is solved.

The problem studied here is related to the \emph{dedicated Relay Broadcast Channel} (dedicated RBC) 
introduced and studied in~\cite{Liang:05z}, and some of the channels
in~\cite{BehboodiPiantanida:13p},\cite{HuWangMaWu:20p}.
Note that the model solved in~\cite[Thm. 40]{Liang:05z} refers to a broadcast channel where one user is degraded
with respect to the relay, and the other is reversely degraded, hence it differs considerably from our problem.

\section{Problem Formulation}
\label{sec:problem_formulation}
\begin{figure}
{\scalefont{0.75}
\begin{tikzpicture}[auto,node distance = 0.8cm and 0.8cm,
  encoder/.style={rectangle,draw=black,inner sep=2pt,minimum height=0.6cm,minimum width=1.2cm},
      channel/.style={rectangle,draw=black,inner sep=2pt,minimum height=0.6cm,minimum width=1.4cm},
      decoder/.style={rectangle,draw=black,inner sep=2pt,minimum height=0.6cm,minimum width=1.2cm},
      helper/.style={ellipse,draw=black,text=black,fill=cyan!20,minimum height=0.6cm,minimum width=1.2cm}]
      \node (m) {};
      \node (m_decoded) {};
      \node [encoder] (encoder) [right=of m] {Encoder};
      \node [channel]  (channel) [right=of encoder] {Channel $P$};
      \node [decoder]  (decoder) [right = 1.2cm of channel] {Decoder};
      \node [channel] (relay) [above = 1.1cm of channel] {Relay};
       \node (m_decoded) [right=of decoder] {};
             \node [inner sep=0] (exitY1) at ($(channel.north)+(-0.3,0)$) {};
              \node [inner sep=0] (inputX1) at ($(channel.north)+(0.3,0)$) {};
                 \node [inner sep=0] (inputY1) at ($(relay.south)+(-0.3,0)$) {};
                \node [inner sep=0] (exitX1) at ($(relay.south)+(0.3,0)$) {};
                \node [helper] (helper) [above right = 0.9cm and 1.2cm of channel] {Helper};
       \tikzstyle{every path}=[draw,->]
        \draw [->] (encoder.east) to node {$X$} (channel.west);
        \path (channel.east) to node {$Y$} (decoder.west);
        \path (m) -- node[above,pos=0.4]{$m$}(encoder);
         \path (decoder) to node {$\hat{m}$} (m_decoded);
         \path (exitY1) to node {$Y_1$} (inputY1);
          \path (exitX1) to node {$X_1$} (inputX1);
          \path (relay.east) to node {$X_2$} (helper);
            \draw [->,black] (helper) to node {$Y_2$} (decoder.north);
\end{tikzpicture}
}
\caption{The relay channel $\tilde{P}$, composed of the channel $P$ and a helper, modeled as an
additional channel $P(y_2|x_2)$ of capacity $C_1$. 
The helper channel $P(y_2|x_2)$ and the main RC $P(y,y_1|x,x_1)$ are decoupled.  }
\label{fig:RC_h}
\end{figure}
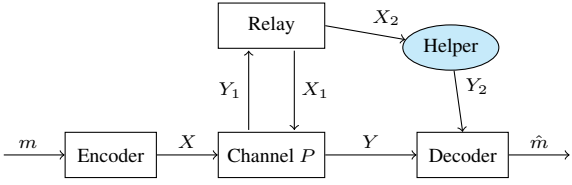

A discrete memoryless relay channel is a quintuple
$\{{\cal X},{\cal X}_1,P(y,y_1|x,x_1),{\cal Y}_1,{\cal Y}\}$, where 
${\cal X}$, ${\cal X}_1$, ${\cal Y}_1$ and ${\cal Y}$ are the alphabets of
the channel input, relay input, relay output, and channel output respectively. 
All alphabets are assumed finite. In the sequel we refer to the RC
as $P(y,y_1|x,x_1)$ or just $P$, when the model is understood from the context. 
An RC is said to be \emph{degraded} if we can write
\be
P(y,y_1|x,x_1) = P(y|x_1,y_1)P(y_1|x,x_1).\label{eq:deg_def_1}
\ee
The capacity of the degraded RC is given by (\cite{CoverElGamal:79p,ElGamalKim:11b}):
\be
C=\max_{p(x,x_1)}\min\{I(X,X_1;Y),I(X;Y_1|X_1)\}.
\label{eq:drc_capacity}
\ee
An RC $P$ with a $C_1$ helper is a channel $\tilde{P}$ defined as
\be
\tilde{P}(y,y_1,y_2|x,x_1,x_2) = P(y,y_1|x,x_1)P(y_2|x_2) ,
\label{eq:helper_def}
\ee
where $P(y_2|x_2)$ is a memoryless channel with capacity $C_1$, from  the relay to the receiver,
decoupled from the RC $P$. Its input and output alphabets are denoted by ${\cal X}_2$ and ${\cal Y}_2$, respectively. 
The RC $\tilde{P}$, depicted in Fig.~\ref{fig:RC_h},
 has an extended relay input $\tilde{X}_1=(X_1,X_2)$ and an extended channel output
$\tilde{Y}=(Y,Y_2)$.
It is easy to verify that if $P(y,y_1|x,x_1)$ is a degraded RC, so is $\tilde{P}$.
The terms that appear in~(\ref{eq:drc_capacity})
\begin{subequations}
\label{eq:primitive_vs_channel}
\begin{IEEEeqnarray}{rCl}
I(X;Y_1|\tilde{X}_1) &=& I(X;Y_1|X_1,X_2) \label{eq:primitive_vs_channel1}\\
I(X,\tilde{X}_1;\tilde{Y}) &= &I(X,X_1,X_2;Y,Y_2),\label{eq:primitive_vs_channel2}
\end{IEEEeqnarray}
\end{subequations}
are both maximized when $X_2$ is capacity achieving for the channel
$P(y_2|x_2)$ and independent of $(X,X_1,Y,Y_1)$.
By~(\ref{eq:drc_capacity}) and~(\ref{eq:primitive_vs_channel}), 
the capacity $\tilde{C}$ of $\tilde{P}$ is given by
\be
\tilde{C} = \max_{p(x,x_1)}\min\{I(X,X_1;Y)+C_1,I(X;Y_1|X_1)\}.
\label{eq:drch_capacity}
\ee
 Hence, $\tilde{C}$ depends on the channel $p(y_2|x_2)$ only via its capacity $C_1$.
 This has been observed before for the primitive relay channel
 (see e.g.~\cite[Sec. 16.7.3]{ElGamalKim:11b}). It is demonstrated here for the case where
 there is also a classical relay in $\tilde{P}$. Therefore the helper can be viewed as a memoryless channel
 $p(y_2|x_2)$ or, equivalently, an additional primitive link of capacity $C_1$.
It is understood that the relay sends encoded messages via the 
input $X_2$. The helper receives these messages and forwards them to the destination via 
$Y_2$. Thus the helper acts as a primitive link, where the burden of coding is put solely on the relay.

\begin{figure}
{\scalefont{0.75}
\begin{tikzpicture}[auto,node distance = 0.8cm and 0.8cm,
  encoder/.style={rectangle,draw=black,inner sep=2pt,minimum height=0.6cm,minimum width=1.2cm},
      channel/.style={rectangle,draw=black,inner sep=2pt,minimum height=0.6cm,minimum width=1.4cm},
      decoder/.style={rectangle,draw=black,inner sep=2pt,minimum height=0.6cm,minimum width=1.2cm},
      helper/.style={ellipse,draw=black,text=black,fill=cyan!20,minimum height=0.6cm,minimum width=1.2cm}]
      \node (m) {};
      \node (m_decoded) {};
      \node [encoder] (encoder) [right=of m] {Encoder};
      \node [channel]  (channel) [right=of encoder] {Channel $P$};
      \node [decoder]  (decoder) [right = 1.2cm of channel] {Decoder};
      \node [channel] (relay) [above = 1.1cm of channel] {Relay};
       \node (m_decoded) [right=of decoder] {};
             \node [inner sep=0] (exitY1) at ($(channel.north)+(-0.3,0)$) {};
              \node [inner sep=0] (inputX1) at ($(channel.north)+(0.3,0)$) {};
                 \node [inner sep=0] (inputY1) at ($(relay.south)+(-0.3,0)$) {};
                \node [inner sep=0] (exitX1) at ($(relay.south)+(0.3,0)$) {};
                \node [helper] (helper) [above right = 0.9cm and 1.2cm of channel] {Helper};
       \tikzstyle{every path}=[draw,->]
        \draw [->] (encoder.east) to node {$X$} (channel.west);
        \path (channel.east) to node {$Y$} (decoder.west);
        \path (m) -- node[above,pos=0.3]{$m,m'$}(encoder);
         \path (decoder) -- node[above,pos=0.6] {$\hat{m},{\color{red}\hat{m}'}$} (m_decoded);
         \path (exitY1) to node {$Y_1$} (inputY1);
          \path (exitX1) to node {$X_1$} (inputX1);
          \path (relay.east) to node {$X_2$} (helper);
            \draw [->,red] (helper) to node {$Y_2$} (decoder.north);
\end{tikzpicture}
}
\caption{The relay channel with unreliable helper. The message $m$ is always decoded.
The message $m'$ is decoded only if the signal $Y_2$ arrives to the decoder. }
\label{fig:RC_uh}
\end{figure}
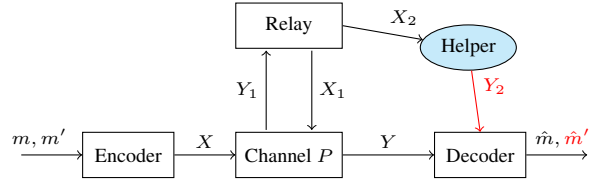

Assume now that the helper is not reliable. 
That is, the encoder and relay do not know what is the actual channel - 
they only know that it is either $P$ or $\tilde{P}$. 
We proceed to describe the workings of a code for this scenario.
Fix a transmission length $n$, and integers $\nu$, $\nu'$. Define two sets of messages
${\cal N}=[1:\nu]$ and ${\cal N'}=[1:\nu']$. 
Messages $m\in{\cal N}$
are always decoded at the destination, whether or not the helper is active.
To decode the messages $m'\in {\cal N}$, the decoder needs the signal $Y_2$,
coming from the helper.  The encoder maps the pair $(m,m')$ to the codeword
${\bml x}(m,m')$ and sends it via the channel. Based on the signal 
${\bml y}_{1}$ he receives, the relay sends ${\bml x}_1$ and ${\bml x}_2$, in a strictly causal manner.
At the destination, the message $m$ is always decoded, whereas $m'$ is decoded if ${\bml y}_2$ is present. 
The model is depicted in Fig.~\ref{fig:RC_uh}.
A formal definition of a code for this setup is given next.

\begin{definition}
\label{def:RC_uh_code}
An $(n,\nu,\nu',\epsilon)$ code for the RC $P$ with unreliable helper $P(y_2|x_2)$
is an encoder
\[
f:\ {\cal N}\times{\cal N}'\rightarrow {\cal X}^n
\]
a causal relay encoder
\begin{subequations}
\label{eq:relay_enc}
\be
\label{eq:relay_enc_vec}
\bml{g} = (g_1,g_2\ldots g_n)
\ee
where
\be
g_i:{\cal Y}_1^{i-1}\rightarrow {\cal X}_{1,i}\times{\cal X}_{2,i},
\label{eq:relay_enc_i}
\ee
\end{subequations}
\label{eq:dec}
and a pair of channel decoders
\begin{subequations}
\begin{IEEEeqnarray}{rCl}
& & \phi: {\cal Y}^n\rightarrow{\cal N}\label{eq:dec_y} \\
& & \phi': {\cal Y}^n\times{\cal Y}_2^n \rightarrow {\cal N}\times{\cal N}'
\label{eq:dec_yy1}
\end{IEEEeqnarray}
\end{subequations}
such that 
\begin{subequations}
\label{eq:prob_err_def}
\begin{IEEEeqnarray}{rCl}
P_e &=& \frac{1}{\nu\nu'}\sum_{m,m'} P(S_{m} | f(m,m'),\bml{g}(\bml{y}_1))\leq\epsilon
\label{eq:pe_def}\\
P_{e'} &=& \frac{1}{\nu\nu'}\sum_{m,m'} P(S'_{m,m'} | f(m,m'),\bml{g}(\bml{y}_1))\leq\epsilon
\label{eq:pe_prime_def}
\end{IEEEeqnarray}
where
\begin{IEEEeqnarray}{rCl}
S_{m} &=& \left\{\bml{y}:\ \phi(\bml{y})\not= m\right\}\label{eq:sm_def}\\
S'_{m,m'} &=& \left\{(\bml{y},\bml{y}_2):\ \phi'(\bml{y},\bml{y}_2)\not= (m,m')\right\}.
\label{eq:sm_prime_def}
\end{IEEEeqnarray}
\end{subequations}
\end{definition}
The rates of the code are given by
\[
R=\frac{\log \nu}{n},\quad R'=\frac{\log \nu'}{n}.
\]
A rate pair $(R,R')$ is said to be achievable
 if for any $\epsilon>0$ and sufficiently large $n$ there 
exists an $(n,2^{nR},2^{nR'},\epsilon)$ code for $P$
with unreliable helper $P(y_2|x_2)$. The capacity region ${\cal C}$
of the channel $P$ with unreliable helper $P(y_2|x_2)$ is the closure
of the set of achievable rates $(R,R')$. 

\begin{remark}
\label{remark:1}
By Definition~\ref{def:RC_uh_code}, if all pairs $(R,R')$ satisfying
\begin{subequations}
\label{eq:remark1}
\begin{IEEEeqnarray}{rCl}
R&<&a \label{eq:remark1a}\\
R'&<& a' \label{eq:remark1b}
\end{IEEEeqnarray}
are achievable, then so are the pairs
\begin{IEEEeqnarray}{rCl}
\tilde{R}&<&a \label{eq:remark1c}\\
\tilde{R}+\tilde{R}'&<& a+a'. \label{eq:remark1d}
\end{IEEEeqnarray}
\end{subequations}
Indeed, note that the decoder~$\phi'$ 
in~(\ref{eq:dec_yy1}) has access to $Y^n$, which consists of all 
the information that decoder $\phi$ has.
Thus we can ignore part of the bits that $\phi$ decodes, 
and attribute them to~$\phi'$.
\end{remark}

\section{Main Result}
\label{sec:main}
Let ${\cal R}(C_1)$ be the set of all pairs $(R,R')$ satisfying
\begin{subequations}
\label{eq:def_set_R}
\begin{IEEEeqnarray}{rCl}
R &\leq& I(U,X_1;Y) \label{eq:def_set_R_R}\\
R+R' &\leq& \min\left\{I(X,X_1;Y)+C_1,I(X;Y_1|X_1),\right.\nonumber\\
          &      &         \left. I(U,X_1;Y) + I(X;Y_1|U,X_1)\right\}
\label{eq:def_set_R_RpRprime}
\end{IEEEeqnarray}
for some $P(u,x,x_1)$ such that 
\be
U\markov(X,X_1)\markov(Y,Y_1).\label{eq:set_R_markov}
\ee
\end{subequations}
 Our main result is stated next.
\begin{theorem}
\label{theo:main1}
For any degraded relay channel $P$ with unreliable helper $P(y_2|x_2)$ 
\[
{\cal C} = {\cal R}(C_1),
\]
where $C_1$ is the capacity of $P(y_2|x_2)$, i.e.,
\[
C_1 = \max_{P_{X_2}} I(X_2;Y_2).
\]
Moreover, to exhaust ${\cal R}(C_1)$,  it is enough
to choose $U$ with alphabet size
$|{\cal U}|\leq |{\cal X}|  |{\cal X}_1| + 2$.
\end{theorem}

\noindent
The proof of Theorem~\ref{theo:main1} is given in Section~\ref{sec:proof}.

\noindent
{\bf Discussion}: Let us examine extreme cases of Theorem~\ref{theo:main1}. 
\begin{enumerate}
\item 
$C_1=0$. Observe that 
\begin{IEEEeqnarray}{cCl}
\lefteqn{\max_{P(u|x,x_1)} I(U,X_1;Y) = I(X,X_1;Y)}\nonumber\\
&\leq& \max_{P(u|x,x_1)} I(U,X_1;Y) + I(X;Y_1|U,X_1)\nonumber
\end{IEEEeqnarray}
hence we can drop the bound on $R$ in~(\ref{eq:def_set_R_R}) and the term
$I(UX_1;Y)+I(X;Y_1|UX_1)$ in~(\ref{eq:def_set_R_RpRprime}) and obtain
\[
R+R' \leq\max_{P(x,x_1)}\min\{I(X,X_1;Y),I(X;Y_1|X_1)\},
\]
 the capacity of the channel without helper. The form of
this bound reflects the fact that the bits that are decodable without the helper can be
arbitrarily allocated between $R$ and $R'$, decoded by $\phi$ and $\phi'$,
respectively.
\item 
$R=0$, or don't care. Here we try to maximize the rate with the helper.
We get for $R'$ the capacity of $\tilde{P}$, eq.~(\ref{eq:drch_capacity}).
For details, see the discussion on $U=X_1$ below.
\item
$R'=0$. Here we aim at maximizing the rate when the helper is absent,
whether or not $C_1>0$. In this case~(\ref{eq:def_set_R}) reduce to~(\ref{eq:drc_capacity}).
The technical details are omitted.
\end{enumerate}
We next examine extreme choices of $U$. For $U=X$ we obtain the capacity of $P$,
the channel without helper. With $U=X_1$ we have
\begin{subequations}
\label{subeq:check}
\begin{IEEEeqnarray}{rCl}
R &< &I(X_1;Y)\label{subeq:check_a}\\
R+R' &<& \min\{ I(X,X_1;Y)+C_1,I(X;Y_1|X_1)\}. \label{subeq:check_b}
\end{IEEEeqnarray}
\end{subequations}
Note that~(\ref{subeq:check_b}) coincides with~(\ref{eq:drch_capacity}).
Therefore, with a properly designed coding scheme, we can guarantee
that when the helper is present we get the maximal rate $\tilde{C}$ (eq.~(\ref{eq:drch_capacity})),
but can still decode $nI(X_1;Y)$ bits when it is absent. 
Observe that substituting a null random variable for $U$ leads to the same result.

\section{The Gaussian Case}
\label{sec:Gaussian_example}
In this section we derive the capacity region of the Gaussian degraded RC with unreliable helper. 
Adding input constraints and passing to continuous alphabets can be done with
the same arguments used for the MAC, BC and relay channels in~\cite{ElGamalKim:11b},
thus we omit these details here.

\begin{figure}
{\scalefont{0.75}
\begin{tikzpicture}[auto,node distance = 0.8cm and 0.8cm,
  encoder/.style={rectangle,draw=black,inner sep=2pt,minimum height=0.6cm,minimum width=0.7cm},
      channel/.style={rectangle,draw=black,inner sep=2pt,minimum height=0.6cm,minimum width=0.9cm},
      decoder/.style={rectangle,draw=black,inner sep=2pt,minimum height=0.6cm,minimum width=0.7cm},
      adder/.style={circle,draw=black,inner sep=2pt,minimum size=0.3cm},
      helper/.style={ellipse,draw=black,text=black,fill=cyan!20,minimum height=0.6cm,minimum width=1.0cm}]
      \node (m) {};
      \node (m_decoded) {};
      \node [encoder] (encoder) [right=of m] {Enc.};
      \node [adder]  (chan_adder) at ($(encoder.east)+(0.7,0.0)$) {$+$};
       \node [adder]  (chan_adder2) at ($(chan_adder.east)+(2.4,0.0)$) {$+$};
      \node  (Z_noise) [below=of chan_adder] {};
       \node  (Z1_noise) [below=of chan_adder2] {};
       \node (m_decoded) [right=of decoder] {};
             \node [inner sep=0] (exitY1) at ($(chan_adder.east)+(0.3,0)$) {};
             \node [channel] (relay) at ($(exitY1)+(1.1,1.5)$) {Relay};
                  \node [decoder]  (decoder) [right = 0.8cm of chan_adder2] {Dec.};
              \node [inner sep=0] (inputX1) at ($(chan_adder.north)+(0.2,0)$) {};
                 \node [inner sep=0] (inputY1_pre) at ($(relay.west)+(-0.2,0)$) {};
                \node [inner sep=0] (exitX1) at ($(relay.east)+(0.0,-0.15)$) {};
                \node [inner sep=0.0] (exitX1_pre) at ($(relay.east)+(0.2,-0.15)$) {};
                \node [inner sep=0] (exitX2) at ($(relay.east)+(0.0,0.15)$) {};
                 \node [inner sep=0] (exitX2_pre) at ($(relay.east)+(0.2,0.15)$) {};
                \node [helper] (helper) [below right = -0.4cm and 1.1cm of relay] {Helper};
       \tikzstyle{every path}=[draw,->]
        \draw [->] (encoder.east) to node {$X$} (chan_adder.west);
        \draw [->] (exitY1.center) to node {$Y_1$} (inputY1_pre.center) -- (relay.west);
        \path (chan_adder.east) -- (chan_adder2.west);
       \path (chan_adder2.east) to node {$Y$} (decoder.west);
        \path (Z_noise) to node {$Z\sim N(0,\sigma^2_z)$} (chan_adder);
        \path (Z1_noise) to node {$Z_1\sim N(0,\sigma^2_1)$} (chan_adder2);
        \path (m) -- node[above,pos=0.3]{$m,m'$}(encoder);
         \path (decoder) -- node[above,pos=0.6] {$\hat{m},{\color{red}\hat{m}'}$} (m_decoded);
           \path (exitX1.center) -- (exitX1_pre.center) to node {$X_1$} (chan_adder2.north);
          \path (exitX2.center)  -- (exitX2_pre.center) to node {$X_2$} (helper);
            \draw [->,red] (helper) to node {$Y_2$} (decoder.north);
\end{tikzpicture}
}
\caption{The Gaussian degraded relay channel with unreliable helper. }
\label{fig:RC_uh_Gaussian}
\end{figure}
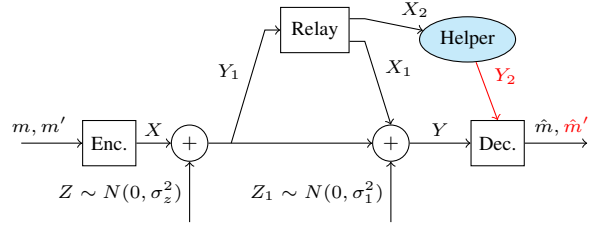

\begin{figure}
    \centering
    \includegraphics[width=\linewidth]{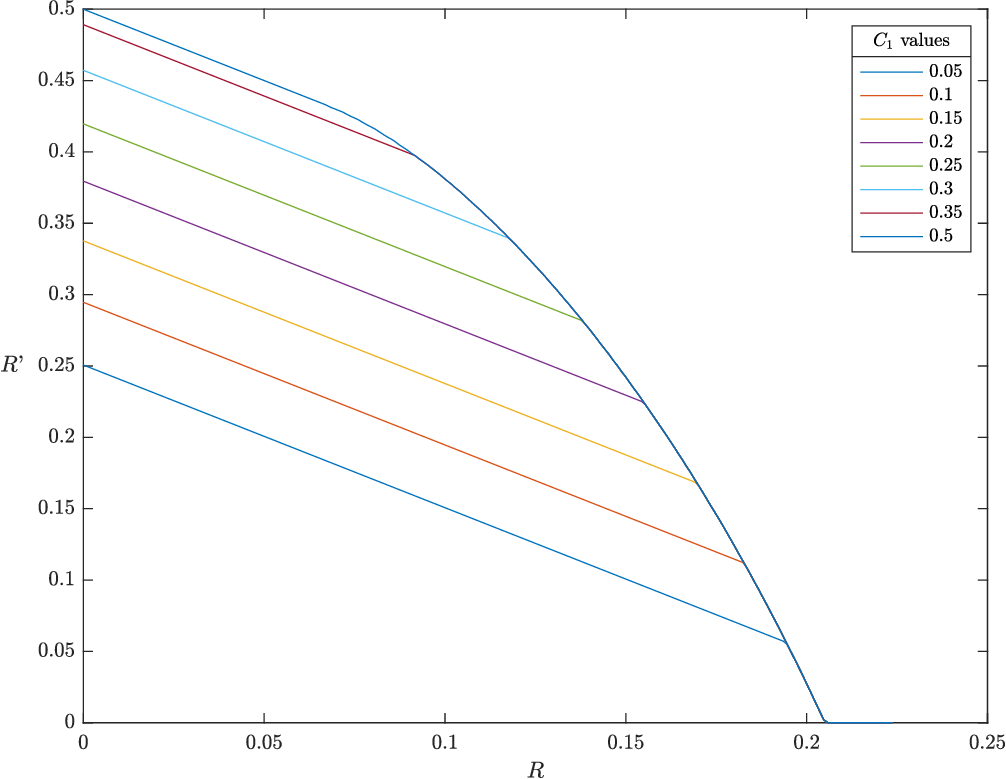}
    \caption{The capacity region of the Gaussian degraded relay channel with unreliable helper, 
    for $P=P_1=\sigma_z^2=1$,
    $\sigma_1^2=10$.}
    \label{fig:cap_Gaussian}
\end{figure}

The channel model is given by
\begin{subequations}
\label{subeq:Gaussian_example}
\begin{IEEEeqnarray}{rCl}
Y_1&=& X+Z\label{subeq:Gaussian_example_Y1}\\
Y  &=& Y_1+X_1+Z_1 = X+X_1+Z+Z_1\label{subeq:Gaussian_example_Y}
\end{IEEEeqnarray}
where $Z$, $Z_1$ are independent Gaussian noises
\be
Z\sim N(0,\sigma_z^2),\ \ \ \  Z_1\sim N(0,\sigma_1^2)\label{subeq:Gaussian_example_noise_power}
\ee
and the transmitter and relay are subject to power constraints
\be
\EE(X^2)\leq P,\ \ \ \EE(X_1^2)\leq P_1.\label{subeq:Gaussian_example_input_constraints}
\ee
\end{subequations}
The channel is depicted in Fig.~\ref{fig:RC_uh_Gaussian}.
Let ${\cal R}(P,P_1,C_1)$ be the set of all pairs $(R,R')$ satisfying
\begin{subequations}
\label{subeq:Gaussian_region_A}
\begin{IEEEeqnarray}{rCl}
R &\leq& C\left(\frac{\olsi{\alpha\beta}P+P_1+2\sqrt{\olsi{\alpha}P P_1}}{\alpha\beta P +\sigma_z^2+\sigma_1^2}\right)
                  \label{subeq:Gaussian_region_A_R}\\
R+R' &\leq& \min\left\{ C\left(\frac{P+P_1+2\sqrt{\olsi{\alpha}P P_1}}{\sigma_z^2+\sigma_1^2} \right) + C_1,
                          C\left(\frac{\alpha P}{\sigma_z^2} \right), \right.\nonumber\\
           & &              \left. C\left(\frac{\olsi{\alpha\beta}P+P_1+2\sqrt{\olsi{\alpha}P P_1}}{\alpha\beta P +\sigma_z^2+\sigma_1^2}\right)
                          +  C\left(\frac{\alpha\beta P}{\sigma_z^2} \right) \right\}\nonumber\\
            & &
                   \label{subeq:Gaussian_region_A_RpRp}       
\end{IEEEeqnarray}
\end{subequations}
for some $\alpha,\beta\in[0,1]$, where $\olsi{\alpha}=1-\alpha$ and $C$ is the capacity function:
\[
C(x) = \half\log(1+x).
\]
Our main result for the Gaussian channel is given next
\begin{theorem}
\label{theo:cap_Gaussian}
The capacity region of the Gaussian degraded relay channel with unreliable helper of capacity $C_1$
is  given by
\[
{\cal C}(P,P_1) = {\cal R}(P,P_1,C_1).
\]
\end{theorem}
\noindent
The proof of Theorem~\ref{theo:cap_Gaussian} is given in Section~\ref{sec:proof_cap_Gaussian}.
Fig.~\ref{fig:cap_Gaussian} depicts the capacity region 
for a few values of the helper capacity $C_1$.

\section{Proof of Theorem~\ref{theo:main1}}
\label{sec:proof}
\subsection{Converse part of Theorem~\ref{theo:main1}}
\label{subsec:proof_converse}
We start with a sequence of  $(n,2^{nR},2^{nR'},\tilde{\epsilon}_n)_{n\geq 1}$ codes,
with $\limn\tilde{\epsilon}_n=0$. By Fano inequality
\begin{IEEEeqnarray}{rCl}
\lefteqn{n(R-\epsilon_n) \leq I(m;Y^n) = \sum_{i=1}^n I(m;Y_i|Y^{i-1})} \nonumber\\
                     &\leq& \sum_{i=1}^n I(m,Y^{i-1};Y_i)
                          \leq \sum_{i=1}^n I(m,Y^{i-1},X_{1,i};Y_i) \label{eq:Fano_R}
\end{IEEEeqnarray}
where $\limn\epsilon_n = 0$. Applying Fano inequality on the sum rate,
we get
\begin{IEEEeqnarray}{rCl}
\lefteqn{n(R+R'-\epsilon_n) \leq I(m,m';Y^n,Y_2^n)}\nonumber\\
                      & = & I(m,m';Y^n)+I(m,m';Y_2^n|Y^n) \nonumber\\
                         &\leq& I(m,m';Y^n)+I(m,m',X_2^n,Y^n;Y_2^n)\nonumber\\
                         &\stackrel{(a)}{=}&  I(m,m';Y^n)+I(X_2^n;Y_2^n)\nonumber\\
                          &\stackrel{(b)}{\leq}&  I(m,m';Y^n)+nC_1\nonumber\\
                          &=& \sum_{i=1}^n I(m,m';Y_i|Y^{i-1}) + nC_1\nonumber\\
                          &\leq& \sum_{i=1}^n I(m,m',X_i,X_{1,i};Y_i|Y^{i-1}) + nC_1\nonumber\\ 
                           &\leq& \sum_{i=1}^n I(m,m',X_i,X_{1,i},Y^{i-1};Y_i) + nC_1\nonumber\\ 
                             &=& \sum_{i=1}^n I(X_i,X_{1,i};Y_i) + nC_1,
                             \label{eq:sum_rate_1}
\end{IEEEeqnarray}
where $(a)$ holds due to the Markov chain 
\[
(m,m',X^n,X_1^n,Y_1^n,Y^n)\markov X_2^n\markov Y_2^n
\]
and $(b)$ due to the definition of $C_1$ in Theorem~\ref{theo:main1}.
Using $I(m';Y^nY_2^n|m)\leq I(m';Y^nY_1^n|m)$ we also have
\be
n(R+R'+\epsilon_n) \leq I(m;Y^n) +I(m';Y^n,Y_1^n|m).\label{eq:sum_rate_2_b1}
\ee
For the first term in~(\ref{eq:sum_rate_2_b1}) we have, following~(\ref{eq:Fano_R}),
\be
I(m;Y^n)\leq \sum_{i=1}^n I(m,Y^{i-1},X_{1,i};Y_i),\label{eq:sum_rate_2_b2}
\ee
and for the second term in~(\ref{eq:sum_rate_2_b1})
\begin{IEEEeqnarray}{rCl}
\lefteqn{ I(m';Y^n,Y_1^n|m)=}\nonumber\\
             & & \sum_{i=1}^n I(m';Y_i,Y_{1,i}|m,Y^{i-1},Y_1^{i-1})\nonumber\\
            &\stackrel{(c)}{=}&  \sum_{i=1}^n I(m',X_i;Y_i,Y_{1,i}|m,Y^{i-1},Y_1^{i-1},X_{1,i})\nonumber\\
             &\leq& \sum_{i=1}^n I(m',X_i,Y_1^{i-1};Y_i,Y_{1,i}|m,Y^{i-1},X_{1,i})\nonumber\\
            &=&  \sum_{i=1}^n I(X_i;Y_i,Y_{1,i}|m,Y^{i-1},X_{1,i})\label{eq:sum_rate_2}
\end{IEEEeqnarray}
where $(c)$ holds since $X_i$ is a function of $(m,m')$ and $X_{1,i}$ a function of $Y_1^{i-1}$.
The third bound on the sum rate:
\begin{IEEEeqnarray}{rCl}
\lefteqn{n(R+R'-\epsilon_n)\leq (m,m';Y^n,Y_1^n)}\nonumber\\
&=& \sum_{i=1}^n I(m,m';Y_i,Y_{1,i}|Y^{i-1},Y_1^{i-1})\nonumber\\
&\stackrel{(d)}{=}& \sum_{i=1}^n I(m,m';Y_i,Y_{1,i}|Y^{i-1},Y_1^{i-1},X_{1,i})\nonumber\\
&\leq& \sum_{i=1}^n I(m,m',Y^{i-1},Y_1^{i-1};Y_i,Y_{1,i}|X_{1,i})\nonumber\\
&=& \sum_{i=1}^n I(m,m',Y^{i-1},Y_1^{i-1},X_i;Y_i,Y_{1,i}|X_{1,i})\nonumber\\
&=& \sum_{i=1}^n I(X_i;Y_i,Y_{1,i}|X_{1,i})\label{eq:sum_rate_3}
\end{IEEEeqnarray}
 where $(d)$ holds since $X_{1,i}$ is a function of $Y_1^{i-1}$. Define 
 \be
 U_i=(m,Y^{i-1}).\label{eq:def_U_i}
\ee
Collecting~(\ref{eq:Fano_R})-(\ref{eq:sum_rate_3}), we have
\begin{subequations}
\label{subeq:bounds_A}
\begin{IEEEeqnarray}{rCl}
R-\epsilon_n &\leq& \one\sum_{i=1}^n I(U_i,X_{1,i};Y_i)\label{eq:bound_R_i}\\
R+R'-\epsilon_n &\leq& \one\sum_{i=1}^n I(X_i,X_{1,i};Y_i) + C_1    \label{eq:bound_sum1_i}\\
R+R'-\epsilon_n &\leq& \one\sum_{i=1}^n I(U_i,X_{1,i};Y_i)\nonumber\\
                           &  & \mbox{} + \one\sum_{i=1}^n I(X_i;Y_i,Y_{1,i}|U_i,X_{1,i})
\label{eq:bound_sum2_i}\\
R+R'-\epsilon_n &\leq& \one\sum_{i=1}^n I(X_i;Y_i,Y_{1,i}|X_{1,i}).\label{eq:bound_sum3_i}
\end{IEEEeqnarray}
\end{subequations}
Apply now the standard steps - introduce a time sharing random variable
$Q$, uniformly distributed on $[1:n]$. The bounds~(\ref{subeq:bounds_A}) become
\begin{subequations}
\label{subeq:bounds_B}
\begin{IEEEeqnarray}{rCl}
R-\epsilon_n &\leq& I(U_Q,X_Q;Y_Q|Q) \label{eq:bound_R_i_B}\\
R+R'-\epsilon_n &\leq& I(X_Q X_{1,Q};Y_Q|Q) +C_1 \label{eq:bound_sum1_i_B}\\
R+R'-\epsilon_n &\leq& I(U_Q X_{1,Q};Y_Q|Q)\nonumber\\
   & & \mbox{} + I(X_Q;Y_Q Y_{1,Q}|U_Q X_{1,Q} Q) \label{eq:bound_sum2_i_B}\\
R+R'-\epsilon_n &\leq& I(X_Q;Y_Q Y_{1,Q}|X_{1,Q}Q).   
\end{IEEEeqnarray}
\end{subequations}
Define  the random variables
\begin{IEEEeqnarray}{rCl}
\tilde{U}=U_Q,\quad & &U = (\tilde{U},Q),\quad X=X_Q,\quad X_1=X_{1,Q}\nonumber\\
 & & Y=Y_Q,\quad Y_1=Y_{1,Q},\label{eq:UXX1YY1_def}
\end{IEEEeqnarray}
observe that the conditional distribution of $(Y,Y_1)$ given $(X,X_1)$ is our
original channel, and that the following inequalities hold
\begin{subequations}
\label{subeq:time_sharing_Q}
\begin{IEEEeqnarray}{rCl}
I(\tilde{U},X_1;Y|Q) &\leq& I(\tilde{U},Q,X_1;Y)\nonumber\\
        & & \mbox{} = I(U,X_1;Y)\label{eq:time_sharing_Q_1}\\
I(X,X_1;Y|Q) &\leq& I(X,X_1;Y) \label{eq:time_sharing_Q_2}\\
I(X;Y,Y_1|X_1,Q) &\leq& I(X,Q;Y,Y_1|X_1) \nonumber\\
     & &   = I(X;Y,Y_1|X_1).\label{eq:time_sharing_Q_3}
\end{IEEEeqnarray}
\end{subequations}
Using~(\ref{eq:UXX1YY1_def}) and~(\ref{subeq:time_sharing_Q}) in~(\ref{subeq:bounds_B})
we obtain
\begin{subequations}
\label{eq:outer}
\begin{IEEEeqnarray}{rCl}
R-\epsilon_n &\leq& I(U,X_1;Y) \label{eq:outer_R}\\
R+R' -\epsilon_n &\leq& \min\left\{I(X,X_1;Y)+C_1,I(X;Y,Y_1|X_1),\right.\nonumber\\
          &      &         \left. I(U,X_1;Y) + I(X;Y,Y_1|U,X_1)\right\}.
\label{eq:outer_R_Rp}
\end{IEEEeqnarray}
\end{subequations}
The Markov chain~(\ref{eq:set_R_markov}) holds by the channel definition.
We proceed next to bound the alphabet size of $U$. 
Let
\begin{equation}
L=|{\cal X}| |{\cal X}_1| +1 \label{eq:L_def}
\end{equation}
and define the $L$ functions
\begin{IEEEeqnarray}{rl}
P_{XX_1|U}(x,x_1|u) & \quad \mbox{$L-2$ functions}\label{eq:Lminus2_functions}\\
H(Y|X_1,U=u) & \nonumber\\
H(Y_1|X_1,U=u)\nonumber
\end{IEEEeqnarray} 
By the Support Lemma~\cite[Appendix C]{ElGamalKim:11b},
\cite[Lemma 3.3.4]{CsiszarKorner:82b}, 
there exists a random variable $U'$ with alphabet 
\begin{equation}
|{\cal U}'|\leq L\label{eq:U_size}
\end{equation}
such that $P_{XX_1}$ is preserved, and
\begin{subequations}
\label{subeq:I1_I2_values}
\begin{IEEEeqnarray}{rCl}
I(U',X_1;Y) &=& I(U,X_1;Y)\label{eq:I1_value}\\
I(X;Y,Y_1|U',X_1) &=& I(X;Y,Y_1|U,X_1). \label{eq:I2_value}
\end{IEEEeqnarray}
\end{subequations}
Due to~(\ref{eq:U_size}), the r.h.s. of~(\ref{eq:outer}) does not depend on $n$,
and we can take the limit $n\rightarrow\infty$, thus obtaining
\begin{subequations}
\label{eq:outer_B}
\begin{IEEEeqnarray}{rCl}
R &\leq& I(U,X_1;Y) \label{eq:outer_R_B}\\
R+R'  &\leq& \min\left\{I(X,X_1;Y)+C_1,I(X;Y,Y_1|X_1),\right.\nonumber\\
          &      &         \left. I(U,X_1;Y) + I(X;Y,Y_1|U,X_1)\right\}.
\label{eq:outer_R_Rp_B}
\end{IEEEeqnarray}
\end{subequations}
Note that thus far we have not used the assumption that the RC $P$ is degraded,
so~(\ref{eq:outer_B}) forms an outer bound for any RC with unreliable helper.
By~(\ref{eq:deg_def_1}) and~(\ref{eq:set_R_markov}) we have
\begin{IEEEeqnarray}{rCl}
I(X;Y,Y_1|X_1)&=&I(X;Y_1|X_1)\label{eq:deg1}\\
I(X;Y,Y_1|U,X_1)&=&I(X;Y_1|U,X_1),\label{eq:deg2}
\end{IEEEeqnarray}
concluding the proof of  the converse.
\qed

\subsection{Direct part of Theorem~\ref{theo:main1}}
\label{subsec:proof_direct}
The proof of the direct part in based on random coding
and $\delta$-typicality decoding. We use the definitions, notation
and $\delta$-convention of~\cite{CsiszarKorner:82b}.
To save space, the definitions are omitted.
In the sequel, block-Markov coding and binning are employed, 
where bins of $m$ are sent via $X_1$,
and bins of the extra message $m'$ are sent via the helper input $X_2$.
Since the helper is not guaranteed to deliver the bin number, superposition coding
is used at the encoder, where the cloud center is always decoded at the destination,
and the satellite message is decoded only if the helper does send the bin number.
Note that superposition coding is employed only at the encoder, 
whereas the relay has one layer of codewords.

\subsubsection{Code construction}
\label{subsubsec:code_construction}
Pick a joint distribution $P(u,x,x_1)$. 
Fix the rates $R$, $R'$, two real numbers $R_2<R$, $R_2'<R'$,
and choose a large integer $B$. We construct $B$ codebooks ${\cal C}_b$,
$b\in[1:B]$, each of length $n$. Each codebook ${\cal C}_b$ is generated randomly, 
independently of the other codebooks, as follows.
\begin{itemize}
\item Generate $2^{nR_2}$ $n$-length codewords $\bml{x}_1^{(b)}(l)$,
$l\in[1:2^{nR_2}]$, iid according to $P(x_1)$.
\item For each $l\in[1:2^{nR_2}]$, generate $2^{nR}$ $n$-length codewords 
$\bml{u}^{(b)}(m|l)$, $m\in[1:2^{nR}]$, independently according to
$\prod_{i=1}^n P(u_i|x_{1,i}^{(b)}(l)).$
\item For each pair $\bml{x}_1^{(b)}(l)$, $\bml{u}^{(b)}(m|l)$,
$l\in[1:2^{nR_2}]$, $m\in[1:2^{nR}]$, generate $2^{nR'}$ $n$-length codewords
$\bml{x}^{(b)}(m'|m,l)$, $m'\in[1:2^{nR'}]$ independently, according to
\[
\prod_{i=1}^n P(x_i| u_i^{(b)}(m|l), x_{1,i}^{(b)}(l)).
\]
\end{itemize}
The code ${\cal C}_b$ is the collection
\begin{IEEEeqnarray}{rCl}
\lefteqn{   \left\{ \bml{x}_1^{(b)}(l),\bml{u}^{(b)}(m|l), \bml{x}^{(b)}(m'|m,l), \right.   } \nonumber\\
 & & \left. l\in[1:2^{nR_2}],\quad m\in[1:2^{nR}],\quad m'\in[1,2^{nR'}] \right\}.
 \label{eq:C_b}
\end{IEEEeqnarray}
Partition the set of messages $m\in[1:2^{nR}]$ into $2^{nR_2}$ bins, where bin $l$ contains messages
\be
m\in\left[(l-1)2^{n(R-R_2)}+1:l2^{n(R-R_2)}\right].\label{eq:bin_m}
\ee
Similarly, 
partition the set of messages $m'\in[1:2^{nR'}]$ into $2^{nR'_2}$ bins, where bin $l'$ contains messages
\be
m'\in\left[(l'-1)2^{n(R'-R'_2)}+1:l'2^{n(R'-R'_2)}\right].\label{eq:bin_m'}
\ee
Reveal the codebooks ${\cal C}_b$ and the partitions to the encoder, relay, and decoder.

Before transmission begins, the sender chooses independently $2(B-1)$ messages
$m_1,m_2\ldots,m_{B-1},\ \ m_b\in[1:2^{nR}]$, and
$m'_1,m'_2\ldots,m'_{B-1},\ \ m'_b\in[1:2^{nR'}]$.
Set $m_0=m'_0=m_B=m'_B=1$. Denote by $l_b$ (resp. $l'_b$) the bin in which 
$m_b$ (res. $m'_b$) resides. The coding technique involves detection
of the messages by the relay. Thus for $\hat{m}_b$ (resp. $\hat{m}'_b$) detected by the relay,
we denote by $\hat{l}_b$ (resp. $\hat{l}'_b$) the bin number in which the estimate
$\hat{m}_b$ (resp. $\hat{m}'_b)$ resides. Finally, $\bml{y}^{(b)}$ and $\bml{y}^{(b)}_1$
stand for the $n$-length vectors
received at the destination and relay during block $b$, respectively.

\subsubsection{Coding}
\label{subsubsec:coding} We describe the operations of the main encoder and relay. 
The notation in the tables below follows~\cite{ElGamalKim:11b}.

\noindent
{\bf Encoder}: In block $b$, the encoder sends
$\bml{x}^{(b)}(m'_b|m_b,l_{b-1})$.

\noindent
{\bf Relay}: When transmission starts ($b=1$), the relay encoder sends
$\bml{x}_1^{(1)}(l_0)$.
At the beginning of block $b$ ($b\geq 2$), the relay looks for $\hat{m}_{b-1},\hat{m}'_{b-1}$
such that
\begin{IEEEeqnarray}{cCl}
\left( \bml{x}^{(b-1)}(\hat{m}'_{b-1}|\hat{m}_{b-1},\hat{l}_{b-2}),\bml{y}_1^{(b-1)} \right) \nonumber\\
        \in {\cal T}_{X,Y_1|X_1}(\bml{x}_1^{(b-1)}(\hat{l}_{b-2})) \label{eq:relay_dec_1}
\end{IEEEeqnarray}
where $\hat{l}_{b-2}$ is the number of bin in which $\hat{m}_{b-2}$  resides.
Note that $l_0$ is known since $m_0$ is fixed a priori.
Table~\ref{tab:relay_decoding} lists the decoding error events and the resulting rate constraints,
under the assumption that decoding of $m_{b-2}$ was correct (hence the conditioning
on $X_1$ in the joint PMFs).
\begin{table}[h!]
    \begin{center}
    \caption{Decoding error events at the relay}
    \label{tab:relay_decoding}
    \begin{tabular}{c|c|c}
    $(\hat{m},\hat{m}')$ & Joint PMF & Rate constraint\\
    \hline
    (1,e) & $P_{U|X_1} P_{X|UX_1} P_{Y_1|UX_1} $ &  $R'< I(X;Y_1|UX_1)$\\
    (e,1) &  $ P_{X|X_1} P_{Y_1|X_1}$      &  $R< I(X;Y_1|X_1)$\\
    (e,e) & $  P_{X|X_1} P_{Y_1|X_1}$      &  $R + R'< I(X;Y_1|X_1)$\\
    \end{tabular}
  \end{center}
 \end{table}

Therefore for the relay decoding to succeed, we need
\begin{subequations}
\label{eq:relay_dec_conditions}
\begin{IEEEeqnarray}{rCl}
R' &<& I(X;Y_1|U,X_1) \label{eq:relay_dec_condition_1}\\
R+R' &<& I(X;Y_1|X_1). \label{eq:relay_dec_condition_2}
\end{IEEEeqnarray}
\end{subequations}
During block $b$, the relay sends $\bml{x}^{(b)}_1(\hat{l}_{b-1})$ via the channel,
and the bin number $\hat{l}'_{b-1}$ via the helper link of capacity $C_1$. For this we need
\be
R_2'<C_1.\label{eq:relay_tx_helper}
\ee

\subsubsection{Decoding at the destination}
\label{subsubsec:decoding_destination}
We distinguish between the case where the signal $Y_2$ is present (i.e., the helper is active),
and the case where it is absent.

\noindent
{\bf Helper absent}: At the end of block $b$, the decoder looks for $\hathat{l}_{b-1}$ such that
\be
\left(\bml{x}^{(b)}_1(\hathat{l}_{b-1}),\bml{y}^{(b)}\right) \in {\cal T}_{X_1,Y}.\label{eq:dec_helper_absent_1}
\ee
For this step to succeed, we need
\be
R_2< I(X_1;Y). \label{eq:dec_helper_absent_R2}
\ee
Then he looks in bin $\hathat{l}_{b-1}$ for a message $\hathat{m}_{b-1}$ such that
\be
\left(\bml{u}^{(b-1)}(\hathat{m}_{b-1}|\hathat{l}_{b-2}),\bml{y} \right) 
\in {\cal T}_{UY|X_1}\left(\bml{x}^{(b-1)}_1(\hathat{l}_{b-2})\right) \label{eq:dec_helper_absent_3}
\ee
which requires
\be
R-R_2< I(U;Y|X_1).\label{eq:dec_helper_absent_4}
\ee
Collecting~(\ref{eq:dec_helper_absent_R2}) and~(\ref{eq:dec_helper_absent_4}) we obtain
the rate bound
\be
R<I(U;Y|X_1)+I(X_1;Y) = I(U,X_1;Y).\label{eq:dec_helper_absent_5}
\ee

\noindent
{\bf Helper present}: At the end of block $b$, the decoder performs joint decoding of $(m_{b-1},m'_{b-1})$.
He first decodes $\hathat{l}_{b-1}$ as in the case where the helper is absent,
resulting in the rate constraint~(\ref{eq:dec_helper_absent_R2}).
He then looks in bin $\hathat{l}_{b-1}$ and bin $l'_{b-1}$ (which was sent by the relay via the helper)
for a pair $(\hathat{m}_{b-1},\hathat{m}'_{b-1})$ such that
\begin{IEEEeqnarray}{cCl}
\left( \bml{u}^{(b-1)}(\hathat{m}_{b-1}|\hathat{l}_{b-2}), 
\bml{x}^{(b-1)}(\hathat{m}'_{b-1}|\hathat{m}_{b-1},\hathat{l}_{b-2}), \bml{y}^{(b-1)}\right) \nonumber\\
 \in {\cal T}_{UXY|X_1}(\bml{x}^{(b-1)}_1(\hathat{l}_{b-2})). \label{eq:dec_helper_present_1}
\end{IEEEeqnarray}
Table~\ref{tab:destination_decoding} lists the decoding error events at the destination, 
and the resulting rate constraints,
under the assumption that decoding of $m_{b-2}$ and $l_{b-1}$ were correct.
\begin{table}[h!]
    \begin{center}
    \caption{Decoding error events at the destination}
    \label{tab:destination_decoding}
    \begin{tabular}{c|c|c}
    $(\hathat{m},\hathat{m}')$ & Joint PMF & Rate constraint\\
    \hline
    (1,e) & $ P_{X|UX_1} P_{Y|UX_1} $ &  $\begin{array}{ccc} R'-C_1 \\
                                                                                                       < I(X;Y|UX_1) \end{array} $\\
\hline
    (e,1) &  $ P_{UX|X_1} P_{Y|X_1}$      &  $ \begin{array}{ccc} R-R_2 \\
                                                                                                      < I(UX;Y|X_1) \end{array}$\\
\hline                                                                                                      
    (e,e) & $  P_{UX|X_1} P_{Y|X_1}$      &  $ \begin{array}{ccc} R-R_2 + R'-C_1 \\
                                                                                                      < I(UX;Y|X_1) \end{array}$\\
\hline                                                                                                      
    \end{tabular}
  \end{center}
 \end{table}
 Table~\ref{tab:Tx_Unreliable_helper} summarizes the transmission and decoding steps
 of our scheme, for the encoder, relay, and destination.
 
\begin{table*}[h]
   \centering
   \caption{Transmission and decoding table for the relay channel with unreliable helper}
   \label{tab:Tx_Unreliable_helper}
   \begin{IEEEeqnarraybox}[\IEEEeqnarraystrutmode%
     ]{c"c"c"c"c"c}
     \toprule
     b & 1 & 2 & 3 &  \ldots & B \\
     \midrule
     X & \bml{x}^{(1)}(m'_1|m_1,l_0) &  \bml{x}^{(2)}(m'_2|m_2,l_1) &  \bml{x}^{(3)}(m'_3|m_3,l_2) &\cdots 
                             &  \bml{x}^{(B)}(1|1,l_{B-1})\\ 
   \midrule     
    \begin{array}{cc} \mbox{Relay dec.}\\
                               \mbox{at block end} \end{array} & \hat{m}_1,\hat{m}_1' &   \hat{m}_2,\hat{m}_2' &  \hat{m}_3,\hat{m}_3' & & \\
      \midrule
      X_1 & \bml{x}_1^{(1)}(l_0)   & \bml{x}_1^{(2)}(\hat{l}_1)    & \bml{x}_1^{(3)}(\hat{l}_2) & \cdots &  \bml{x}_1^{(B)}(\hat{l}_{B-1})\\
    \midrule
      \mbox{Helper} & - & \hat{l}_1'& \hat{l}_2' &  \cdots & \hat{l}_{B-1}' \\
      \midrule
     \begin{array}{cc} \mbox{$Y$ dec.}\\
                               \mbox{at block end}\\
                               \mbox{if helper absent}  \end{array} & - & \hathat{m}_1 &\hathat{m}_2 & \cdots & \hathat{m}_{B-1}\\
      \midrule
       \begin{array}{cc} \mbox{$Y$ dec.}\\
                               \mbox{at block end}\\
                               \mbox{if helper present}  \end{array} & - & \hathat{m}_1,\hathat{m}_1' &\hathat{m}_2,\hathat{m}_2' & \cdots 
                               & \hathat{m}_{B-1},\hathat{m}_{B-1}'\\
     \bottomrule
   \end{IEEEeqnarraybox}
 \end{table*}

Using~(\ref{eq:dec_helper_absent_R2}) in the rate constraints
 of Table~\ref{tab:destination_decoding}, we obtain
 \begin{subequations}
 \label{eq:dec_helper_present_2}
 \begin{IEEEeqnarray}{rCl}
 R' &<& I(X;Y|U,X_1)+C_1\label{eq:dec_helper_a}\\
 R &<& I(U,X;Y|X_1) + R_2\nonumber\\
     & < & I(U,X;Y|X_1) + I(X;Y)\nonumber\\
     & = & I(X,X_1;Y)  \label{eq:dec_helper_b}\\
 R+R' &<& I(U,X;Y|X_1)+R_2+C_1\nonumber\\
                     &<& I(X,X_1;Y) + C_1. \label{eq:dec_helper_c}
 \end{IEEEeqnarray}
 \end{subequations}
 We collect now the rate constraints imposed by the relay~(\ref{eq:relay_dec_conditions})
 and the main decoder (eq.~(\ref{eq:dec_helper_absent_5}) when the helper is absent, 
 and~(\ref{eq:dec_helper_present_2}) when it is present). Note that~(\ref{eq:dec_helper_absent_5})
 dominates~(\ref{eq:dec_helper_b}), hence
 \begin{subequations}
 \label{subeq:achievable_rates_1}
 \begin{IEEEeqnarray}{rCl}
 R &<& I(U,X_1;Y)\label{eq:achievable_rates_1a}\\
 R' &<& \min\{I(X;Y|UX_1) + C_1, I(X;Y_1|UX_1)\}\label{eq:achievable_rates_1b}\\
 R+R' &<& \min\{ I(X,X_1;Y)+C_1, I(X;Y_1|X_1)\}. \label{eq:achievable_rates_1c}
 \end{IEEEeqnarray}
 \end{subequations}
Due to Remark~\ref{remark:1}, we can replace~(\ref{eq:achievable_rates_1b})
by the sum of~(\ref{eq:achievable_rates_1a}) and~(\ref{eq:achievable_rates_1b}),
resulting in~(\ref{eq:def_set_R_R}),~(\ref{eq:def_set_R_RpRprime}). 
This completes the proof of the direct part.
\qed

\section{Proof of Theorem~\ref{theo:cap_Gaussian}}
\label{sec:proof_cap_Gaussian}
We have to evaluate the region~(\ref{eq:def_set_R}), 
under the input constraints
\be
\EE(X^2)\leq P,\ \ \ \EE(X_1^2)\leq P_1.\label{eq:Gaussian_proof_input_constraints}
\ee
\subsection{Direct Part}
\label{subsec:direct_cap_Gaussian}
Let $X_1$, $U$ and $V$ be independent Gaussian RVs with
\begin{subequations}
\begin{IEEEeqnarray}{rCl}
X_1\sim N(0,P_1)\label{subeq:X1_Gaussian}\\
U\sim N(0,\alpha\olsi{\beta} P)\label{subeq:U_Gaussian}\\
 V\sim N(0,\alpha\beta P) \label{subeq:V_Gaussian}
\end{IEEEeqnarray}
and define $X$ as
\be
X= \sqrt{\olsi{\alpha}\frac{P}{P_1}} X_1 +U +V. \label{eq:def_X1_Gaussian}
\ee
\end{subequations}
The pair $(X,X_1)$ is independent of $(Z,Z_1)$, and
\be
X+X_1\sim N(0,P+P_1+2\sqrt{\olsi{\alpha}P/P_1}).\label{eq:X_plus_X1}
\ee
Therefore
\begin{IEEEeqnarray}{rCl}
h(Y)&=&h(X+X_1+Z+Z_1) \label{eq:h1} \\
        &=& \half\log\left[2\pi e\left(P+P_1+2\sqrt{\olsi{\alpha}P P_1}+\sigma_z^2+\sigma_1^2\right)\right].
        \nonumber
\end{IEEEeqnarray}
Similarly, differential entropies comprising the region~(\ref{eq:def_set_R}) satisfy
\begin{IEEEeqnarray}{rCl}
\lefteqn{h(Y|U,X_1)}\label{eq:h2}\\
         &=&h(X+X_1+Z+Z_1|UX_1) \nonumber\\
         &=& h\left(\left(1+\sqrt{\olsi{\alpha}P/P_1}\right)X_1 + U +V +Z +Z_1|UX_1\right)\nonumber\\
        &=& h(V+Z+Z_1) = \half\log\left[2\pi e\left(\alpha\beta P+\sigma_z^2+\sigma_1^2\right)\right],
        \nonumber
\end{IEEEeqnarray}
\begin{IEEEeqnarray}{rCl}
h(Y|XX_1) &=& h((X+X_1+Z+Z_1|XX_1)\label{eq:h3} \\
                  & = & h(Z+Z_1) = \half\log[2\pi e(\sigma_z^2+\sigma_1^2)]\nonumber
\end{IEEEeqnarray}
\begin{IEEEeqnarray}{rCl}
h(Y_1|X_1) &=& h(X+Z|X_1) = h(U+V+Z)\nonumber\\
                   &=& \half\log[2\pi e(\alpha P + \sigma_z^2)]\label{eq:h4}\\
 h(Y_1|XX_1) &=& \half\log(2\pi e\sigma_z^2)\label{eq:h5}
\end{IEEEeqnarray}
\begin{IEEEeqnarray}{rCl}
h(Y_1|UX_1) &=& h\left(\sqrt{\olsi{\alpha}P/P_1}X_1+U+V+Z|U,X_1\right)\nonumber\\
&=& h(V+Z) = \half\log[2\pi e(\alpha\beta P+\sigma_z^2)]\label{eq:h6}\\
h(Y_1|UXX_1) &=& h(Z) = \half\log(2\pi e\sigma_z^2). \label{eq:h7}
\end{IEEEeqnarray}
Hence
\begin{IEEEeqnarray}{rCl}
I(UX_1;Y) &=& \half\log\left(\frac{P+P_1+2\sqrt{\olsi{\alpha}PP_1} +\sigma_z^2+\sigma_1^2}{\alpha\beta P +\sigma_z^2+\sigma_1^2}\right)
                \nonumber\\
                &=& C\left(\frac{\olsi{\alpha\beta}P+P_1+2\sqrt{\olsi{\alpha}P P_1}}{\alpha\beta P +\sigma_z^2+\sigma_1^2}\right)
\label{eq:I1}\\
I(XX_1;Y) &=& C\left(\frac{P+P_1+2\sqrt{\olsi{\alpha}PP_1}}{\sigma_z^2+\sigma_1^2}\right)\label{eq:I2}\\
I(X;Y_1|X_1) &=& C\left(\frac{\alpha P}{\sigma_z^2}\right)\label{eq:I3}\\
I(X;Y_1|UX_1)&=& C\left(\frac{\alpha\beta P}{\sigma_z^2}\right)\label{eq:I4}
\end{IEEEeqnarray}
where~(\ref{eq:I1}) is due to~(\ref{eq:h1}) and (\ref{eq:h2}), (\ref{eq:I2}) due to~(\ref{eq:h1}) and~(\ref{eq:h3}),
(\ref{eq:I3}) holds by~(\ref{eq:h4}) and~(\ref{eq:h5}), and~(\ref{eq:I4}) by~(\ref{eq:h6}) and~(\ref{eq:h7}).
Substituting~(\ref{eq:I1})-(\ref{eq:I4}) in~(\ref{eq:def_set_R}) yields the achievability result.

\subsection{Converse Part}
\label{subsec:converse_cap_Gaussian}
The proof of the converse makes use of basic linear estimation arguments, and the conditional version
of the entropy power inequality (EPI). This mix is attributed to the structure
of the problem: we have a classical relay part, and a superposition part due to the uncertainty of the helper.
Accordingly, the terms in~(\ref{eq:def_set_R}) that do not contain the cloud center $U$ can be upper bounded 
using arguments similar to those in~\cite[Sec. IV]{CoverElGamal:79p}. 
The terms in~(\ref{eq:def_set_R}) that contain $U$
are bounded using the conditional EPI, 
as usually done in the proof of the converse for the Gaussian broadcast channel~\cite[Sec. 5.5]{ElGamalKim:11b}.
The details are given here for completeness.

Let $U,X,X_1$ be zero mean RVs, with arbitrary joint distribution, independent of $Z,Z_1$.
Define the parameter $\alpha$ as
\be
\alpha = 1 - \rho^2,\nonumber
\ee
here $\rho$ is the correlation coefficient
\be
\rho = \frac{\EE(XX_1)}{\sqrt{PP_1}}.\nonumber
\ee
By Cauchy-Shwartz inequality, $0\leq\alpha\leq 1$. 
In the sequel we will make use of the following elementary inequalities
\begin{subequations}
\label{subeq:Vars}
\begin{IEEEeqnarray}{rCl}
\Var(X+X_1) &=& \EE\left[(X+X_1)^2\right] = P+P_1+2\rho\sqrt{PP_1}\nonumber\\
                     &\leq& P+P_1+2\sqrt{\olsi{\alpha}PP_1}.\label{eq:Var1_bound}
\end{IEEEeqnarray}
Define
\be
\Var[X|X_1] \eqd  \EE\left[(X-\EE(X|X_1))^2|X_1\right], \nonumber
 \label{eq:Var2_def}
\ee
then
\begin{IEEEeqnarray}{rCl}
\EE\Var(X|X_1)  &=&  \EE\left[\left(X-\EE(X|X_1)\right)^2\right]\nonumber\\
                          &\leq& (1-\rho^2)P = \alpha P.\label{eq:Var2_bound}
\end{IEEEeqnarray}
\end{subequations}
The inequality in~(\ref{eq:Var2_bound}) is true because the l.h.s is  the MMSE in estimating $X$ given $X_1$,
and the  r.h.s is the mean square error of the optimal linear estimator~\cite{Hayek:15b}. 
By properties of the differential entropy
\begin{IEEEeqnarray}{rCl}
h(Y) &=& h(X+X_1+Z+Z_1)\nonumber\\
        &\leq&  \half\log\left[2\pi e \Var(X+X_1+Z+Z_1)\right] \nonumber\\
        &=& \half\log\left[2\pi e\left( \Var(X+X_1)+\sigma_z^2+\sigma^2_1\right)\right] \nonumber\\
        &\stackrel{(a)}{\leq}&  \half\log\left[2\pi e\left( P+P_1+2\sqrt{\olsi{\alpha}PP_1}+\sigma_z^2+\sigma^2_1\right)\right],\nonumber\\
     & &    \label{eq:Conv_h1}\\
 h(Y|XX_1) &=& h(X+X_1+Z+Z_1|XX_1) = h(Z+Z_1)\nonumber\\
                   &=& \half\log[2\pi e(\sigma_z^2+\sigma_1^2)], \label{eq:Conv_h2}\\
 h(Y_1|X_1) &=&  h(X+Z|X_1)\nonumber\\
                     &\leq& \half\EE \log\left[2\pi e \Var(X+Z|X_1)\right]\nonumber\\
                     &=& \half\EE \log\left[2\pi e\left(\Var(X|X_1) + \sigma_z^2\right)\right]\nonumber\\
                 &\stackrel{(b)}{\leq}& \half\log\left[2\pi e\left(\EE\Var(X|X_1) +\sigma_z^2\right)\right]\nonumber\\
                 &\stackrel{(c)}{\leq}& \half\log\left[2\pi e\left(\alpha P +\sigma_z^2\right)\right],\label{eq:Conv_h3}\\
h(Y_1|XX_1) &=& h(X+Z|XX_1) = h(Z) = \half\log(2\pi e\sigma_z^2), \label{eq:Conv_h4}
\end{IEEEeqnarray}
where in $(a)$ we used~(\ref{eq:Var1_bound}), in $(b)$ the concavity of the $\log$ function
and in $(c)$ the inequality~(\ref{eq:Var2_bound}). By~(\ref{eq:Conv_h1}) and~(\ref{eq:Conv_h2})
\begin{IEEEeqnarray}{rCl}
I(XX_1;Y) &\leq& C\left(\frac{P+P_1+2\sqrt{\olsi{\alpha}PP_1}}{\sigma_z^2+\sigma_1^2}\right)\label{eq:Conv_I1}
\end{IEEEeqnarray}
and by~(\ref{eq:Conv_h3}) and~(\ref{eq:Conv_h4})
\be
I(X;Y_1|X_1) \leq C\left(\frac{\alpha P}{\sigma_z^2}\right). \label{eq:Conv_I2}
\ee
We proceed to the terms in~(\ref{eq:def_set_R}) that contain $U$
\begin{IEEEeqnarray}{rCl}
h(Y|UX_1) &=& h(X+X_1+Z+Z_1|UX_1)\nonumber\\
                  &\leq& h(X+X_1+Z+Z_1|X_1)\nonumber\\
                  & =& h(X+Z+Z_1|X_1)\nonumber\\
                  &\leq& \half\EE\log\left[2\pi e\Var(X+Z+Z_1|X_1)\right]\nonumber\\
                  &=&  \half\EE\log\left[2\pi e\left(\Var(X|X_1) +\sigma_z^2+\sigma_1^2\right)\right] \nonumber\\
                  &\stackrel{(d)}{\leq}& \half\log\left[2\pi e\left(\alpha P + \sigma_z^2+\sigma_1^2\right)\right]
                  \label{eq:Conv_h5}
\end{IEEEeqnarray}
where inequality $(d)$ holds following $(b)$ and $(c)$ in~(\ref{eq:Conv_h3}). 
On the other hand
\begin{IEEEeqnarray}{rCl}
h(Y|UX_1) &\geq& h(Y|UXX_1) \stackrel{(e)}{=} h(Y|XX_1)\nonumber\\
   &=& \half\log[2\pi e(\sigma_z^2+\sigma_1^2)] \label{eq:Conv_h6}
\end{IEEEeqnarray}
we used the Markov chain $U\markov(X,X_1)\markov(Y,Y_1)$ in $(e)$, 
and the last equality follows~(\ref{eq:Conv_h2}). In view of~(\ref{eq:Conv_h5})
and~(\ref{eq:Conv_h6}), we must have
\be
h(Y|UX_1) = \half\log\left[2\pi e\left(\beta\alpha P + \sigma_z^2+\sigma_1^2\right)\right]
\label{eq:Conv_h7}
\ee
for some $\beta\in[0,1]$. Using~(\ref{eq:Conv_h7}) and~(\ref{eq:Conv_h1}) we obtain the
bound
\begin{IEEEeqnarray}{rCl}
I(UX_1;Y) &\leq& \half\log\left(\frac{P+P_1+2\sqrt{\olsi{\alpha}PP_1}+\sigma_z^2+\sigma_1^2}{\beta\alpha P 
                              +\sigma_z^2+\sigma_1^2}\right)\nonumber\\
                  &=& C\left(\frac{\olsi{\beta\alpha} P +P_1+ 2\sqrt{\olsi{\alpha}PP_1}}{\beta\alpha P+\sigma_z^2+\sigma_1^2}\right).
                  \label{eq:Conv_I3}            
\end{IEEEeqnarray}
It remains to bound $I(X;Y_1|UX_1)$.
Note that
\begin{IEEEeqnarray}{rCl}
h(Y|UX_1) &=& h(Y_1+X_1+Z_1|UX_1)\nonumber\\
                  & = & h(Y_1+Z_1|UX_1),
\label{eq:Conv_h8}
\end{IEEEeqnarray}
where $Y_1$ and $Z_1$ are conditionally independent given $(U,X_1)$.
Hence by the conditional EPI~\cite[Sec. 2.2]{ElGamalKim:11b}
\begin{IEEEeqnarray}{rCl}
2^{2h(Y|UX_1)} &=& 2^{2h(Y_1+Z_1|UX_1)}\nonumber\\
       &\geq& 2^{2h(Y_1|UX_1)} + 2^{2h(Z_1|UX_1)}\nonumber\\
       &=& 2^{2h(Y_1|UX_1)} + 2\pi e \sigma_1^2.\label{eq:Conv_EPI}
\end{IEEEeqnarray}
Substituting~(\ref{eq:Conv_h7}) in~(\ref{eq:Conv_EPI}) we get
\be
h(Y_1|UX_1)\leq \half\log\left[2\pi e(\beta\alpha P +\sigma_z^2)\right].
\label{eq:Conv_h9}
\ee
In addition
\begin{IEEEeqnarray}{rCl}
h(Y_1|UXX_1) &=& h(Y_1|XX_1) = h(Z) \nonumber\\
                     &=& \half\log\left(2\pi e\sigma_z^2\right),
                     \label{eq:Conv_h10} 
\end{IEEEeqnarray}
which, with~(\ref{eq:Conv_h9}) imply
\be
I(X;Y_1|UX_1) \leq C\left(\frac{\beta\alpha P}{\sigma_z^2}\right).\label{eq:Conv_I4}
\ee
Substituting~(\ref{eq:Conv_I1}), (\ref{eq:Conv_I2}), (\ref{eq:Conv_I3}) and (\ref{eq:Conv_I4})
in~(\ref{eq:def_set_R}), we conclude the converse result.

\bibliography{./references_Yossi}{}
\bibliographystyle{IEEEtran}

\end{document}